\documentclass[fleqn,usenatbib]{mnras}

\usepackage{txfonts}

\usepackage[T1]{fontenc}

\DeclareRobustCommand{\VAN}[3]{#2}
\let\VANthebibliography\thebibliography
\def\thebibliography{\DeclareRobustCommand{\VAN}[3]{##3}\VANthebibliography}

\usepackage{graphicx}	

\title[Catsteroseismology]{Catsteroseismology: Survey-based Analysis of Purr-mode Oscillations Suggests Inner Lives of Cats are Unknowable}

\author[Holcomb \& Lam]{
Rae J. Holcomb,$^{1,3}$
Christopher Lam,$^{2,3}$
\\
$^{1}$Department of Physics \& Astronomy, University of California, Irvine, Irvine, CA 92697, USA\\
$^{2}$Department of Astronomy, University of Florida, Gainesville, FL 32607, USA\\
$^{3}$Center for Computational Astrophysics, Flatiron Institute, New York, NY 10010, USA
}

\begin{document}
\label{firstpage}
\pagerange{\pageref{firstpage}--\pageref{lastpage}}
\maketitle

\begin{abstract}

Catsteroseismology, or asterocatsmology, is an unexplored area of observational and theoretical research that proposes to use purr-mode oscillations to study the much-beloved but poorly-understood species \emph{Felis catus}. In this work, we conduct a survey to measure fundamental purrameters of cats and relate them to their purr-modes. Relations between these fundamental cat purrameters, which include physical (eg. size, cuddliness) and personality (eg. aggression, intelligence) traits, and purr-modes can help probe their inner lives and emotions. We find that while purr characteristics tentatively trend with several physical and personality traits, more data is required to better constrain these relationships and infer the direct predictive power of personality on purr-modes, or vice versa. 
\end{abstract}





\section{Introduction}
\label{sec: introduction}

Cats and physicists have had a long-enduring research relationship, dating as far back as the publications of F.D.C. Willard and his human co-authors \citep{willard1975}. In recent years, this fascination has only grown, leading the astrophysics community to begin studies that treat cats as astrophysical objects in the hopes of better understanding our enigmatic feline friends. These have taken the form of deeper investigation of the thought experiments on Schrodinger's Cat \citep{gato2011schroedingers}, treatment of cats as ``floofy" rotational objects \citep{mayorga2021detection}, exoplanet-style detections of P(lan)ET transits \citep{sagynbayeva_first_2022}, and even co-authorship of further manuscripts with cats \citep{armstrong_my_2021}. \footnote{Counterexamples include Wagg, Tzanidakis, Hurtado \& Gilbert-Janziek 2024, which conduct rat-based science.} These efforts have borne significant fruit\footnote{And significant scratches.}, which suggests that there is much more to be learned about cats by applying an astrophysical framework to our studies of them.

Asteroseismology is the study of stellar interiors using their global oscillations, which has allowed for detailed characterization of fundamental stellar purrameters and the probing of stellar interiors. Stars propagate waves in two ways. Pressure modes (also called ``p-modes" or ``acoustic modes") are excited in the convective zone of a star where the restoring force is pressure. Gravity modes (or ``g-modes") originate in the radiative zone, where pressure waves dissipate after short distances and thus the restoring force is buoyancy. Surface oscillations in Sun-like stars are dominated by p-modes.

Analogously, cats also have two primary modes of pulsation, which we dub ``purr-modes" and ``growl-modes". Similarly to p-modes, purr-modes are excited by a convection-like process also known as ``petting", which is comprised of circular motions of a hand or other appendage across the surface of the cat. Appropriately, purr-modes are easily detectable at the surface of the cat. Growl-modes are less well studied, for reasons which we will examine in Section \ref{sec:growl-modes}.

We propose that just as asteroseismology can probe the interior structure of a star, purr-mode and growl-mode oscillations reveal information about the emotional state--that is, the ``interior life"--of cats. Apocryphally, purr-modes are associated with positive emotions within the cat (such as satisfaction, cuddliness, happiness, and affection for their owners) while growl-modes arise from negative emotions (such as aggression, anger, and other ``devilish" inclinations.) However, cats are notoriously inscrutable even to the humans most closely bonded with them, and these connections have yet to be proven by science.

In this paper, we take inspiration from the astrophysical study of stellar pulsations and apply it to cats. Just as asteroseismology probes the interior structure of stars, this new technique, which we dub ``catsteroseismology", aims to draw connections between the oscillatory behavior of cats and their fundamental purrameters, thus shedding light on the inner lives of our dearest feline friends. Section \ref{sec:background} provides background information about the types of pulsations that arise in cats. Section \ref{sec:methods} describes the methods used to collect survey data for this study. Section \ref{sec: results} summarizes our results and discusses possible biases and limitations of this analysis. Our findings are summarized in Section \ref{sec: conclusions}.

\section{Background: Pulsations in Cats and Stars}
\label{sec:background}

\begin{figure}
	\includegraphics[width=\columnwidth]{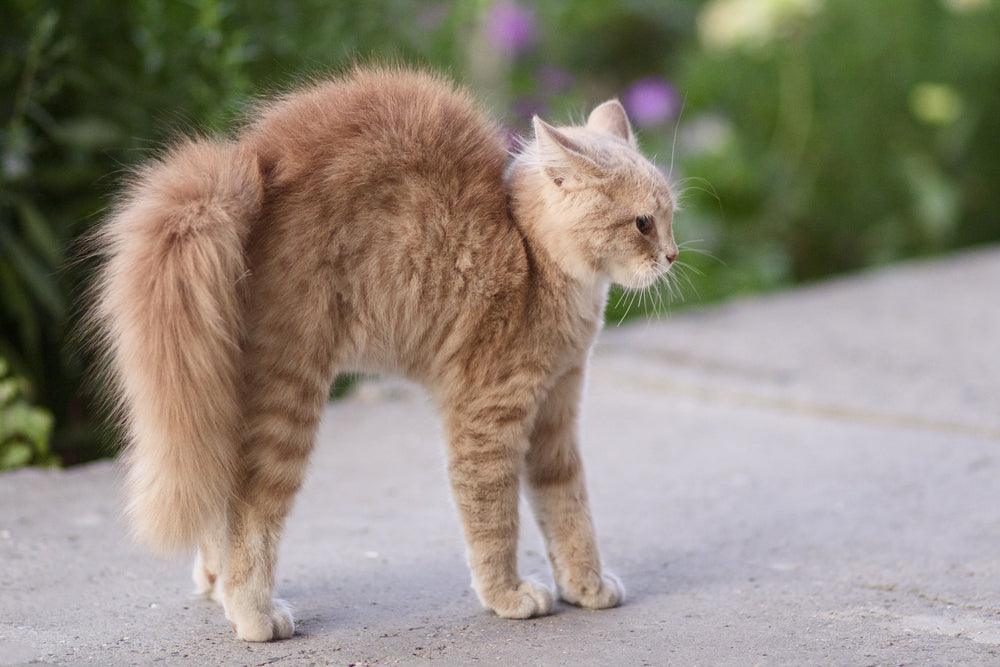}
    \includegraphics[width=\columnwidth]{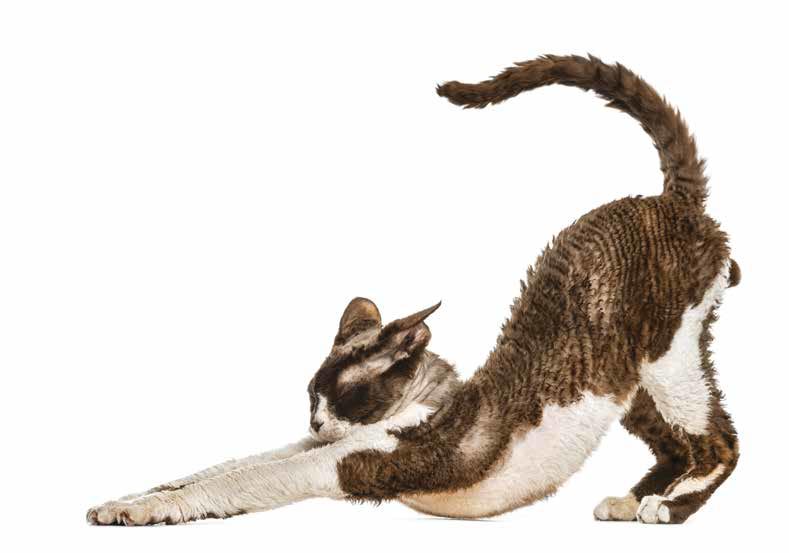}
    \caption{Examples of the two stretch modes: arched-back (source: Adobe stock photo, Standard License) and butt-up (source: Eric Isselee, stock photo). Cats resonate between these two modes at secular timescales, because why would you rush a good stretch?}
    \label{fig:cat-arch-stretch}
\end{figure}

To better understand the inner machinations of cats, we devise a mathematical description of purr-modes. Naturally, we begin with an analog to the basic stellar assumption of hydrostatic equilibrium: we assume that all cats follow the equation for hydrocatic equilibrium, which maintains the familiar form: 
\begin{equation}
    dP = -\rho(P) g(h) dh,
\label{eq:hydrocatic-equilibrium}
\end{equation}
where P is the purr intensity, $\rho$ is the density of the cat\footnote{The mental density, not the physical density usually described as mass per volume.}, g is the growl intensity, and h is the hunger level of the cat. We marginalize over g(h) because growl modes are beyond the scope of this paper (see Section \ref{sec:growl-modes}). The remaining observable is the purr intensity of the cat. When calibrated with appropriate values for each purrameter, we correctly recover typical purr frequencies of 20 to 30 vibrations per second, which is standard for domestic cats \citep{wiki:purr}\footnote{All facts in this and the following paragraph were pulled from the Wikipedia article on ``Purr", and all of which have scientific citations. Most of them even look pretty legit. Unfortunately, the authors of this manuscript couldn't be arsed to hand-copy citations into BibTex format for articles that don't appear in ADS.}.

The mechanism by which cats purr is not definitively known, despite the enduring fascination that humans seem to have with the process \citep{wiki:purr}. Early theories have posited that purring is a hemodynamic process in which sound is produced as the blood runs through the thorax \citep{wiki:purr}. Other theories have focused on the role of the vocal folds within the larynx. Particularly, it is suggestive that all species within Fellidae are capabale of roaring or purring, but not both. This division corresponds definitively with whether that species has a hyoid bone, found in the larynx, that is incompletely (``roarers") or completely (``purrers") ossified \citep{wiki:purr}. Electromyographic studies have suggested that the vocal folds of cats alternately constrict and dilate the glottis rapidly, producing strong harmonics as air vibrates rapidly in the process of exhalation and inhalation \citep{wiki:purr}. The exact mechanisms at play remain unclear. Regardless, it is clear that, like stars, the pulsations of cats reflect their internal structure.\footnote{Thank you, Wikipedia.}

It should also be noted that cats can exhibit other types of acoustic activity than just purr-modes and growl-modes. Longer timescale modulations typically take the form of mechanical distortions, where the cat under goes squashing and stretching but maintains its total volume, as demonstrated in Figure \ref{fig:cat-arch-stretch}. This relates to the fundamental fluid properties of the cat (as we all know, cats are liquids). Cats are susceptible to stochasatic acoustic events, also known as ``meows". Behaviors of these types are adorable, but beyond the scope of this paper.

\begin{figure}
	\includegraphics[width=\columnwidth]{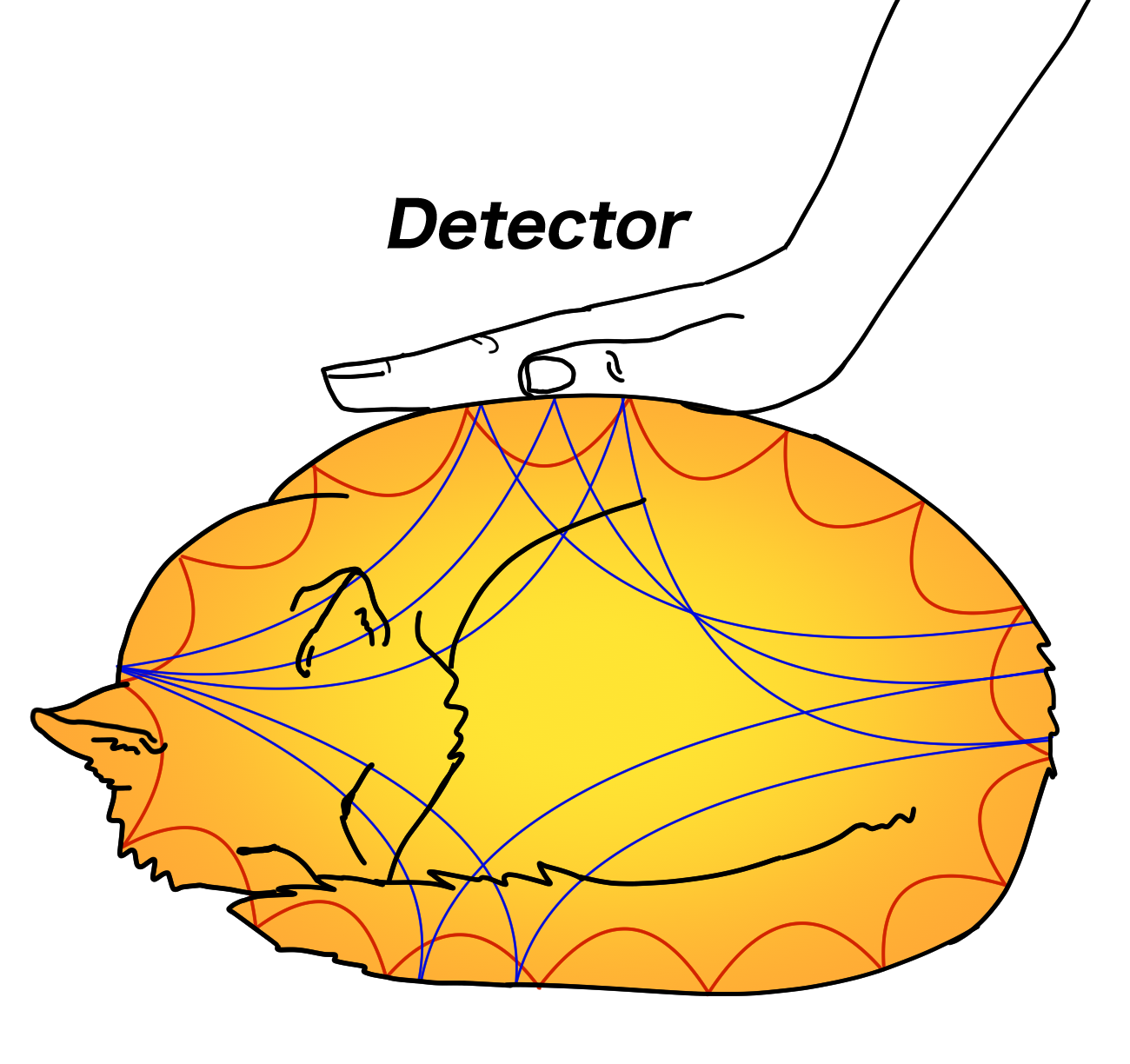}
    \caption{Schematic of a cat producing purr-mode oscillations, and the method by which they may be directly detected.}
    \label{fig:detector}
\end{figure}

\subsection{Purr-Mode Detection Methods}

Pulsations in stars are usually detected by identifying high frequency periodic components in the photometric time series. However, when it comes to cats, there are two potential methods by which we may study their purr-modes. Purring induces high-frequency distortions in the cat. The first method of detecting these is through direct physical contact with the cat, as demonstrated in Figure \ref{fig:detector}. This method is highly effective, but it risks confounding the measurement by inducing a positive feedback loop in which convective ``petting" motions stimulate further purr-modes within the cat. This can be avoided if the researcher practices extreme restraint, and keeps their hand stationary during the period of contact. However, few mortals possess this level of self-discipline.

The second method involves indirect detection. As the cat oscillates, it stimulates sound waves in the surrounding air which can be detected by the human ear. This method has the benefit of not requiring any direct contact with the cat. However, some cats purr very quietly, such that the observing human must place their ear very close to the cat and thus risk disturbing it that way. Ultimately, both methods have their benefits and drawbacks.

\subsection{Growl-Mode Detections \& Issues}
\label{sec:growl-modes}

In Sun-like stars, g-mode oscillations are difficult to observe at the stellar surface and thus most asteroseismic efforts of these stars focus on p-modes. In cats, growl-modes are similarly difficult to study, although for different reasons. The occurrence of growl-modes within a cat is frequently accompanied by outbursts of violent behavior including biting, scratching, and execution of general mayhem, posing significant risk of bodily harm to researchers trying to study them at close quarters (Figure \ref{fig:cat-bite}). Attempts have resulted in numerous injuries. Further pursuit of this line of inquiry has been banned.



\section{Methods: Survey Design}
\label{sec:methods}

We base our investigations on a survey of cat purrameters. To do this, we developed and distributed a survey probing the following observables:

\begin{figure}
	\includegraphics[width=\columnwidth]{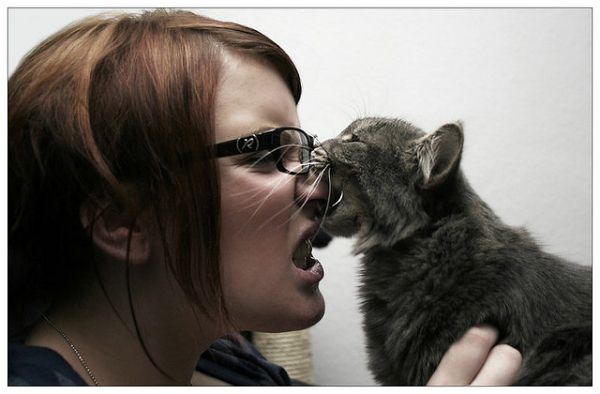}
    \caption{Common failure mode of g-mode oscillation detection attempts (source: pestbarn, CC BY-SA 2.0, via Flickr).}
    \label{fig:cat-bite}
\end{figure}

\begin{table*}
  \centering
    \begin{tabular}{ c|c|c|c|c }
     Observable & Question & Response Type & Min Value & Max Value \\ 
     \hline
     Age & How old is your cat? & Free Response & N/A & N/A \\  
     Mass & How much does your cat weigh in pounds? & Free Response & N/A  & N/A  \\    
     Length & Subjectively, how large is your cat? & 1-10 Scale & Teeny Tiny & Honkin Chonker \\    
     Friendliness & How cuddly is your cat? & 1-10 Scale & Do Not Touch Me Human & Cuddle Bug \\    
     Aggression & How aggressive is your cat? & 1-10 Scale & Not at all & Devourer of Ankles \\    
     Intelligence & How intelligent is your cat? & 1-10 Scale & No thoughts/head empty & Wrote my thesis \\    
     Activity Level & How hyper is your cat? & 1-10 Scale & All day snoozefest & ZOOMIES \\    
     Purr Susceptibility & What percentage of the time will they start to purr when petted? & Percentage & 0 & 100 \\    
     Purr Volume & How loudly does your cat purr? & 1-10 Scale & Barely audible & Foghorn \\      
    \end{tabular}  \caption{Summary of survey questions. Most responses were filled out by human affiliates on behalf of the cats; however, several respondents informed us that the cat was present during survey completion (usually on the keyboard) and thus played a role in their own evaluation.}
  \label{tab:1}
\end{table*}

\begin{itemize}
    \item mass
    \item length
    \item age
    \item friendliness
    \item intelligence
    \item activity level
    \item aggression
    \item purr volume
    \item purr susceptibility
\end{itemize}

Purrameters such as mass and age can be measured absolutely, whereas others require subjective interpretation. Subjective variables were graded by cat owners on a 1-10 scale. The questions and available range of responses for each purrameter can be found in Table \ref{tab:1}. We also asked for qualitative descriptions of the cats' coloration and provided an open response question in which respondents could provide information on "anything else [they] would like to tell us about [their] cat, for the purpose of science".

\section{Results}
\label{sec: results}

\begin{figure}
	\includegraphics[width=\columnwidth]{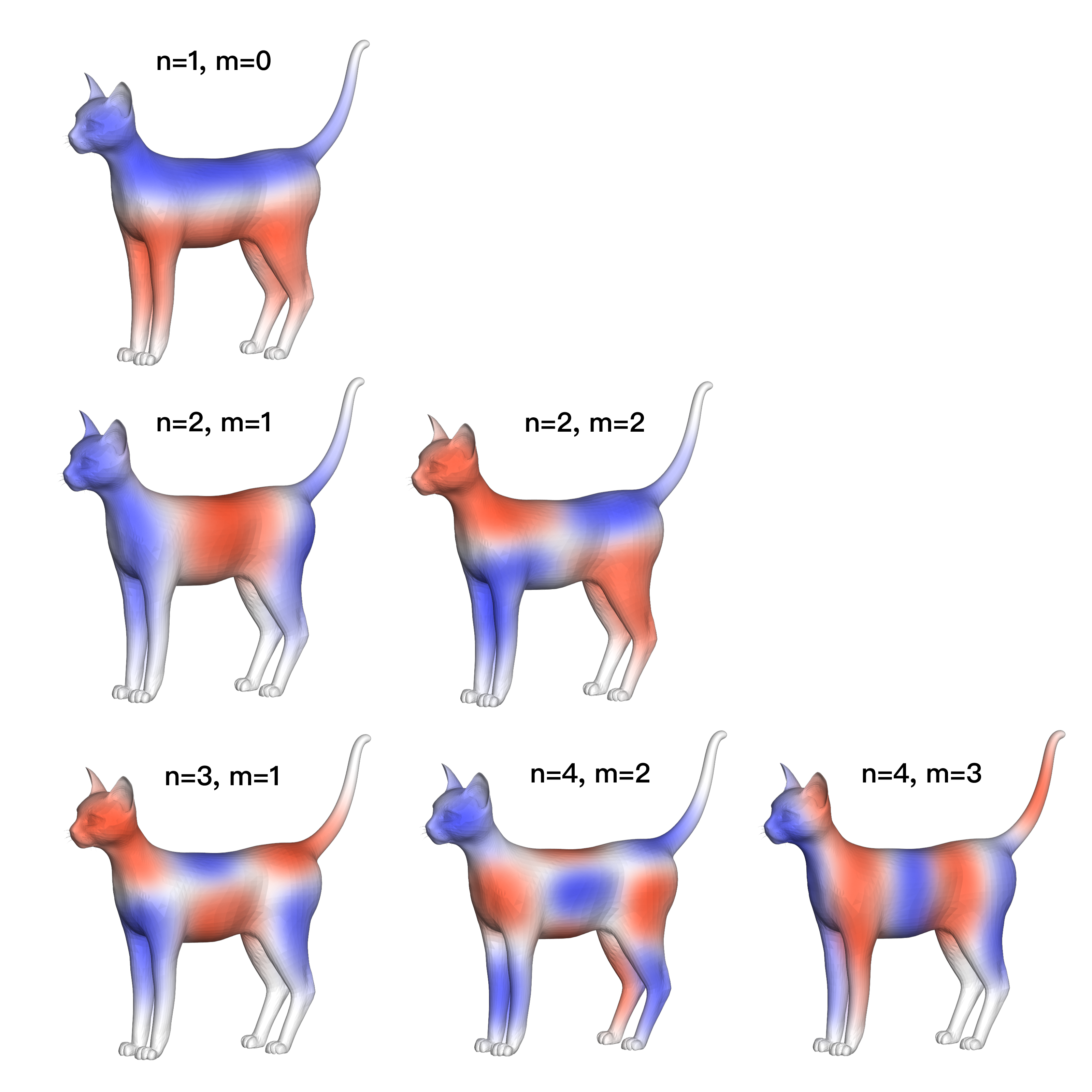}
    \caption{Catsteroseismic modes can be described in the eigenbases of felinical harmonics.}
    \label{fig:cat-modes}
\end{figure}

\begin{figure*}
	\includegraphics[width=\textwidth]{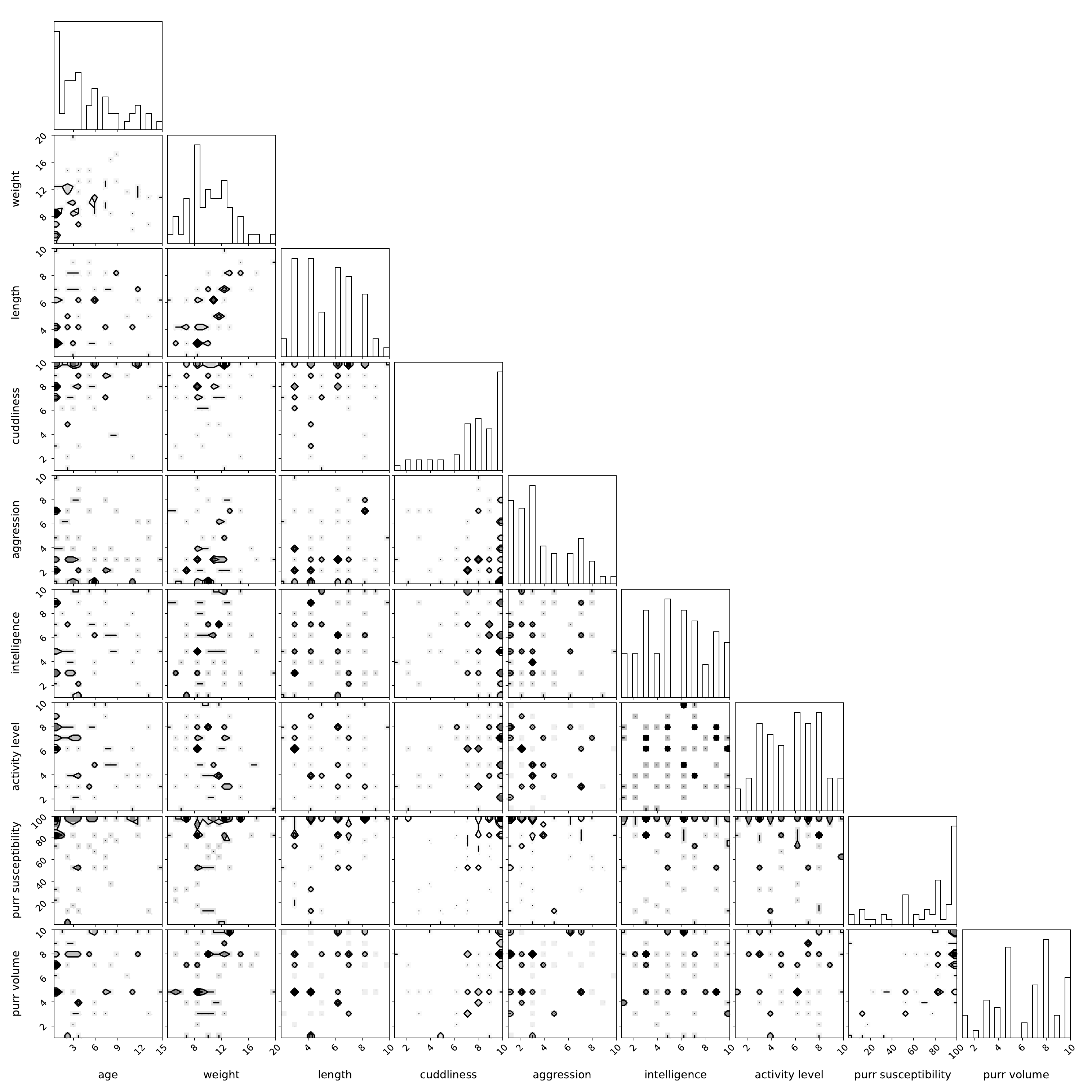}
    \caption{Covariances between the measured purrameters in our survey.}
    \label{fig:corner}
\end{figure*}

Our survey collected data on 145 cats in March 2023. We use the \texttt{corner} software package \citep{Foreman-Mackey2016} to visualize one- and two-dimensional distributions of the data (Figure \ref{fig:corner}).

Interpretation of the one-dimensional data can be made from the histograms in Figure \ref{fig:corner}. We observe that cuddliness and the purr susceptibility are top-heavy distributions. Cats in the sample also tend to be younger and less aggressive. We find as well the following positive covariances: weight and length; age and weight; age and length; weight and aggression; length and aggression; cuddliness and purr volume; length and aggression; age and cuddliness; and aggression and purr volume. We also observe negative covariances between activity level and size, activity level and age, and age and aggression. 

Overall, we find these relationships to be largely self-consistent. For example, the covariance between length and weight and between purr volume and susceptibility are to be expected. We do not find, however, clear trends between the relevant purr-mode quantities (purr volume and susceptibility) and intrinsic personalities. There is a weak association between aggression and purr volume, but the similarly weak association between purr volume and cat size may simply point to larger cats having larger larynxes. Larger cats also tend to purr more often, but they may just be showing off their larger larynxes. 

Most of these covariances are not surprising, but we note that some of these relationships may be a direct function of how they are measured. For example, the greater purr volume among cuddlier cats could be merely a function of how hard a human cuddles a cat. We encourage a separate instrumentation study in which detector arms and detector hands cuddle and pet hundreds of cats at varying strengths and measure their resulting purr volumes\footnote{The authors acknowledge that such a study would require extensive resources to implement. Previous attempts have found that they ``can't hug every cat", and this realization has reduced the researchers to tears \citep{hugeverycat}. We are currently seeking funding to pursue success in this area where so many others have failed.}. Furthermore, the absence of a negative relationship between aggression and purr volume or ease of purr induction (indeed, if anything, these are positive trends) casts doubt on the nigh-universally held belief that purr-modes are associated exclusively with positive emotions. This is supported by the cat Sprout, which ``wants cuddles but is also mad about it".  We strongly encourage further research to consider that cats are highly complex, multi-modal creatures, and that study of their true nature may require consideration in higher dimensional spaces.

We note that intelligence is the only purrameter that does not co-vary with any other purrameter. This suggests two possible interpretations. First, it may that be intelligence truly has no correlation with any other fundamental purrameter. However, cats are complex, with many chaotically interconnected variables required to describe them, and it seems unlikely that something as fundamental as intelligence would be truly independent of all other factors. Therefore, it may also be suggested that this lack of correlation is instead the result of deliberate meddling in the data. Perhaps it would not be unreasonable to consider that sufficiently intelligent cats may be capable of controlling their outward appearance of intelligence (or lack their of) in order to play to human expectations and mask any true underlying signal. With this power, cats would then be able to expertly manipulate their ``owners" into compliant behavior, such as providing food and cuddles when desired. This does not imply that the authors believe in a conspiracy among cats.\footnote{But if the authors disappear to cat jail, the reader will know what happened.}

\subsection{Qualitative survey results}
\label{sec: qualitative}
In order to bring more context to the survey results, we solicited free response comments from humans. These provided a wealth of useful information that would not have been evident from the survey alone. For example, cats feel enmity toward vacuum cleaners, and in one case, rubber bands. Consideration of environmental conditions like these are important for our understanding of cat formation and evolution. 

One human noted that for the first three years of their cat's life, they thought the cat could not purr, even when he was clearly happy. Only after reaching a critical point did the cat begin to purr audibly while being petted or cuddled, albeit quietly. Identifying other cats that exhibit purr reluctance until entering the purring main sequence can inform us not only of the completeness of the survey on purr frequencies, but also hints at distinct evolutionary tracks among cats.

Another striking pattern in these responses was the frequent reports of mayhem perpetrated by these felines. Examples of evil events documented in the survey include unplugging computer monitors for attention (this may also affect survey completeness, since evil cats would therefore be underrepresented if their owners were unable to complete the survey) and the ominously vague ``malicious noncompliance"\footnote{The reader may suppose that this lends further credence to existence of a cat conspiracy. We caution that such lines of thinking are dangerous, and should not be pursued.}. Finally, one human noted that their cat is a ``...gossip girl if I’ve ever seen one. She’s got all the tea and will spill it sitting on your shoulders while you do the dishes." We invite this particular survey-taker to email the authors because a cat spy would provide unprecedented access for future studies.

\subsection{Survey Biases \& Concerns}
\label{section: biases}

\begin{figure}
	\includegraphics[width=\columnwidth]{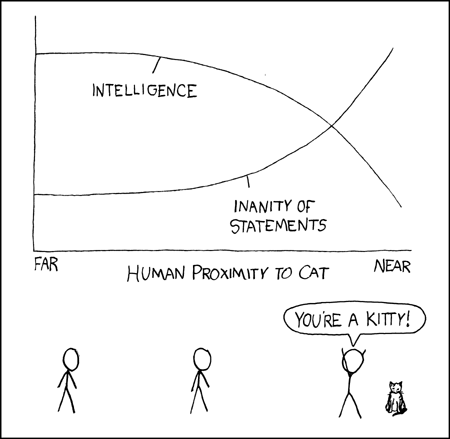}
    \caption{The deleterious effects of cat proximity on the ability of humans to perform objective and rigorous evaluations has been well documented. Reprinted from: https://xkcd.com/231/}
    \label{fig:cat-proximity}
\end{figure}

To participate in the survey, cats required the assistance of their human affiliate, as cats do not have opposable thumbs and cannot type\footnote{This is not strictly true \citep{armstrong_my_2021}; however, instances of typing-capable cats have been the cause of much grief for their human affiliates and we do not encourage such behavior.}. The requirement of this intermediary actor introduces two primary biases. First, it has been well documented that proximity to cats has a strong degradative effect on human mental faculties \citep{xkcd}. An example of this effect is shown in Figure \ref{fig:cat-proximity}. When in close proximity to a cat, humans (particularly the cat's owners) are liable to exhibit regressive behaviors such as ``baby talk" and will make inane statements such as ``Who's a cute little fluffykins?" While it is true that most cats are, in fact, cute little fluffykins, this is likely to compromise our survey respondents' ability to objectively evaluate their cats' purrameters and may introduce a bias towards positive representation of the cats in our survey by their owners.

The second bias effect we anticipate is that most cat owners have high levels of attachment to their cats. This may introduce further biases into certain variables, particularly those with subjective evaluations such as intelligence, aggression, cuddliness, and activity level. As can be seen in Figure \ref{fig:corner}, cat owners preferentially scored their cats as high on cuddliness and low on aggression. However, many survey respondents also reported statements such as ``she is my evil baby", ``He is baby man and a menace", and ``pure evil hidden in a cute body". All the cats mentioned in these quotes were scored significantly higher on cuddliness than on aggression by their owners.

Finally, we note that the demographics of the cat hosts surveyed may be biased. Not all cats belonged to astronomers, but astronomers are overrepresented in the sample. The survey was advertised through word of mouth on various astronomy-related Slack and Discord message boards, with respondents being encouraged to spread the survey further. The majority of our respondents are, therefore, astronomers, though the survey was also distributed among other populations, including a discussion board for moms, an internet forum for webcomic enthusiasts, and the group chat of a roller derby team.


\subsection{Future work}
\label{sec:future-work}

We did not incorporate cat coloration into our analyses in this work; we leave this as future work. Furthermore, from the free response portion of the survey, it is clear that some cats in our survey exist in multi-cat households; however, since we did not directly ask about it, the true cat multiplicity of our sample remains largely unknown. It is possible that cats may experience ``middle-child" or ``single-child" syndrome that affect their purr-mode behaviors. A future survey probing the effect of multiplicity on purr-mode oscillations may reveal distinct populations, which may enable the inference of the architectures of multi-cat systems based on observations of purr-modes, or vice versa. 

We also briefly consider the astrobiological implications of an extension of this survey to study purr-mode oscillations on human moods. Cat purrs generally have a positive effect on the moods of humans. Therefore, it is reasonable to assume that stellar oscillations may contain a similar ``sweet spot" as well. By isolating the optimal mood-boosting frequency range of p-mode oscillations, we can identify planet host stars that oscillate at frequencies most likely to result in happy inhabitants. 

\section{Conclusions}
\label{sec: conclusions}
In this work, we place observational constraints on the relationship between purr-mode oscillations and several cat purrameters, largely divided into personality-based (eg. aggression, intelligence) and tactile-based (eg. cuddliness, size). We find internal self-consistencies (long cats weigh more; cats that purr loudly also purr more often), trends between the tactile and purr-mode purrameters (cuddly cats purr loudly and more often), and trends between the personality and purr-mode purrameters (aggressive cats purr loudly and more often). Physical purr-mode trends, however, may simply be a function of pettable cats being petted more often. Similarly, personality purr-mode trends may be more directly mediated by physical purr-mode trends, such as where aggressive cats, which tend to be larger, purr more loudly. Further study is required to determine whether greater purr volume and frequency is a result of the cat being more assertive, the cat showing off its larger voicebox, or some combination of the two. For now, we are unable to make strong inferences on the relationship between purr-modes and any other meaningful cat property.

Finally, we present this study as a generalizable framework for future study about other demographic cat purrameters, such as how membership in multi-cat systems affects personality, physical properties, and purr-modes. Ultimately we were unable to truly understand the inner lives of cats, but our results have laid the groundwork for further discoveries in the field of catsteroseismology. This endeavor will require collecting vastly more data, which will in turn require the petting of many more cats. We encourage the scientific community to pursue this task with gusto until complete understanding of our feline friends is achieved.


\section*{Acknowledgements}

We would like to thank Daisy, Mavis, Leiko, Socks, Noodle, Penguin, Sibelius, Ruby, Leo, Gizmo, Bellatrix, Carlo, Morgana, Willow, Trouble Puffs, Axel, Atlas, Xander, Kuro, Nils, Creampuff, Phobos, Deimos, Bill, Kira, Alice, Perseus, Percy, Orion, Flash, Remy, Joni, Luna, Rocket, mooney, Daisy, Zodi, Bambam, Izzy, Meeka, Miss, Kitty, Marigold, Kitty, The Entire Nordic Pantheon, Rory, Thalia, Beelzebub, Tikku, Sylvie, Lulu, Harvey, Angel, Patches, Millie, Ocho (aka 'Wub'), Willow, Clementine, Grape, Calvin, Seven, Sylvie, Moonie, Peppa, Pinot, Gritty Kitty, Jefferson, Mabel, Carrot, Frizby, Luna, Honey, Ms. Moneypenny, Kitkat, Bean, Mozzarella, Georgie, Figg, Lazlo, Oona, Syl, George, Coco, Pepper, Charlie, Cecilia, Noli, Ryan, Luna, Jelly, Victor, Amilia, WiggleWorm, Finn, Merry, Pippin, xena, Delilah, Sashimi, Lucky, Wenny, Cookie, Margot, Jasper (aka "bungus, boyo, biggus boigus, bubba"), Squash Blossom, Bear, Jynx, Toast, Tirma, Buster, Ruthie, Mars, Muffin, Bean, Tsar, Ringo, Midi, Porkchop, Pebbles, Alley, Sasha, Leo, Diamonds, Pearls, Meeku, Camper, Hippo, Gizmo!, Medusa, Beethoven, Clara Joy, Jack Jack, Sasha, Salem, Sabrina, Belle, Jasmine, Mei, Leo, Stella, Russell, Sprout, Dester, Opal, Gobi, Cairo, Milo, Divot, Calcifer, Kabu, and their associated humans for generously providing data for this survey. Thank you for deigning to let us live in your homes and worship your feline grace.




\bibliographystyle{mnras}
\bibliography{bib} 



\bsp	
\label{lastpage}
\end{document}